\documentclass[11pt]{article}

\usepackage[T2A]{fontenc}
\usepackage[cp866]{inputenc}
\usepackage{amsthm,amssymb,amsmath}
\usepackage{graphicx}

\usepackage{graphics}

\makeatletter \@addtoreset{equation}{section} \makeatother

\newtheorem{theorem}{Theorem}
\newtheorem{lemma}{Lemma}
\newtheorem{remark}{Remark}

\newtheorem{proposition}{Proposition}

\def\fracd{\displaystyle\frac}
\def\sumd{\displaystyle\sum}
\def\limd{\displaystyle\lim}
\def\intd{\displaystyle\int}
\def\prodd{\displaystyle\prod}

\def\supd{\displaystyle\sup}
\def\E{{\mathbb E}}
\def\C{{\mathbb C}}
\def\R{{\mathbb R}}
\def\1{{\bf 1 }}

\begin{document}
\baselineskip 23pt \vskip 0.5 cm

\title
{EIGENVALUE DISTRIBUTION OF
BIPARTITE LARGE WEIGHTED  RANDOM GRAPHS. RESOLVENT APPROACH}
\author{
$\;$V. Vengerovsky  $\!\!\:\!$, Institute for Low Temperature Physics,
Ukraine}
\date{}
\maketitle
\abstract{
We study eigenvalue distribution of the adjacency
matrix $A^{(N,p, \alpha)}$
of weighted random bipartite graphs $\Gamma= \Gamma_{N,p}$. We assume
that the graphs have
$N$ vertices, the ratio of parts is $\fracd{\alpha}{1-\alpha}$ and  the average number of edges
attached to one vertex  is
$\alpha\cdot p$ or $(1-\alpha)\cdot p$. To each edge of the graph $e_{ij}$ we assign a weight given
by a random variable
$a_{ij}$ with the finite second moment.

We consider the resolvents
$G^{(N,p, \alpha)}(z)$ of $A^{(N,p, \alpha)}$  and study the functions $f_{1,N}(u,z)=\frac{1}{[\alpha N]}\sum_{k=1}^{[\alpha N]}e^{-ua_k^2G_{kk}^{(N,p,\alpha)}(z)}$ and $f_{2,N}(u,z)=\frac{1}{N-[\alpha N]}\sum_{k=[\alpha N]+1}^Ne^{-ua_k^2G_{kk}^{(N,p,\alpha)}(z)}$ in the
limit $N\to \infty$. We derive  closed system of equations that uniquely
determine the  limiting functions $f_{1}(u,z)$ and $f_{2}(u,z)$. This system of
equations allow us to prove the existence of the limiting  measure
 $\sigma_{p, \alpha}$ . The weak convergence in probability  of normalized eigenvalue counting
measures is proved.

\section{Introduction}

The spectral theory of graphs is an actively developing field of
mathematics involving a variety of methods and deep results (see the
monographs \cite{Ch,CDS,GR}).
Given a graph with $N$ vertices, one can associate with it many
different matrices, but the most studied are the adjacency
matrix and the Laplacian matrix of the graph.
Commonly, the set of $N$ eigenvalues of the adjacency matrix
is referred to as the spectrum of the graph.
In these studies, the dimension of the matrix $N$
is usually regarded as a fixed parameter.
The spectra of infinite graphs  is considered
in certain particular cases of graphs having  one or another
regular structure (see for example \cite{JMRR}).

Another large class of graphs, where the limiting transition
$N\to\infty$ provides a natural approximation is represented
by random graphs \cite{B,JLR}. In this branch,
geometrical and topological properties of graphs (e.g. number of connected components, size of
maximal connected component) are  studied for
immense number of random graph ensembles.
One of the classes of the
prime reference is the {\it binomial random graph}
originating by P. Erd\H{o}s (see, e.g. \cite{JLR}).
Given a number $p_N\in (0,1)$, this family of graphs ${\mathbf G}(N,p_N)$
is  defined by taking as $\Omega$ the set of all graphs
on $N$ vertices with the probability
$$
P(G) = p_N^{e(G)} (1-p_N)^{{N \choose 2} - e(G)},
\eqno (1.1)
$$
where $e(G)$ is the number of edges of $G$. Most of the random graphs
studies are devoted to the cases where $p_N\to 0$ as $N\to\infty$.

Intersection of these two branches of the theory of graphs contains
 the spectral theory of random graphs that is still poorly explored.
  However, a number of powerful tools can be employed here, because
 the ensemble of random symmetric $N\times N$ adjacency matrices
  $A_N$ is a  particular
representative of the random matrix theory,
where the limiting transition $N\to\infty$
is intensively studied during half of century since the pioneering
works by E. Wigner \cite{W}.
Initiated by theoretical physics applications,
the spectral theory of random matrices
has revealed deep nontrivial links with many
fields of mathematics.

Spectral properties of random  matrices corresponding to (1.1)
were examined in the limit $N\to\infty$
both in numerical and theoretical physics studies
\cite{E1, E2, E3,MF:91,RB:88,RD:90}.
There are  two major asymptotic regimes:
$p_N \gg 1/N$ and $p_N = O(1/N)$ and corresponding models
can be called the {\it diluted random matrices} and {\it sparse random
matrices}, respectively.
The first studies of
spectral properties of sparse and diluted random matrices
in the physical
literature are related with the works  \cite{RB:88}, \cite{RD:90},
\cite{MF:91}, where equations for the limiting density of states of
sparse random matrices  were
derived. In  papers \cite{MF:91} and \cite{FM:96} a number
of important  results on the universality of the correlation
functions
  and the Anderson localization transition were obtained.
  But all these results were obtained with
   non rigorous  replica method.

On mathematical level of rigour, the  eigenvalue distribution of dilute
random matrices was studied in \cite{KKPS}. It was shown that
  the normalized eigenvalue counting function of
$$
{1\over \sqrt{N p_N} } A_{N,p_N}
\eqno (1.2)
$$
converges in the limit $N, p_N \to \infty$ to the distribution known as the
semicircle or Wigner law
\cite{W}.
The moments of this distribution verify well-known recurrent relation
for the Catalan numbers and can be found explicitly.
Therefore one can say that the dilute random matrices
represent explicitly solvable model (see also \cite{RB:88,RD:90}).

In the series of papers \cite{BG1,BG2,B} and  in \cite{KSV},
the adjacency  matrix and the Laplace matrix of random graphs (1.1)
with $p_N = pN$ were studied.
It was shown that
  this sparse random matrix ensemble can also be viewed as the explicitly
solvable model.

In the present paper we consider a bipartite analogue of  large sparse random  graph. This article
follows the method  of \cite{KSV}. The moment method for this problem was considered in  \cite{V}.


\section{Main results}


    We can introduce the randomly weighted
adjacency matrix of random bipartite graphs.
Let $\Xi=\{a_{ij} ,\; i \!\leq\! j,\;i,j\! \in \!{\mathbb N}\}$ be
the  set of
jointly independent identically distributed (i.i.d.) random
variables determined
  on the same probability space.
  We set $a_{ji}\!=\! a_{ij}$ for $i\!\leq\! j \;$.

Given $ 0\!<p\!\leq \!N$, let us define the family
  $D^{(p)}_N\!=\!\{d^{(N,p)}_{ij},
\; i\!\leq\! j,\; i,j\in \overline{1,N}\}$ of jointly independent
random variables
\begin{equation}
d^{(N,p)}_{ij}\!=\! \left\{ \begin{array}{ll} 1,&
\textrm{with} \ \textrm{probability } \ p/N ,
\\0,& \textrm{with} \ \textrm{probability} \ 1-p/N ,\\ \end{array}
\right.
\end{equation}
We set $d_{ji}= d_{ij}$ and assume that $ D^{(p)}_N$
is independent from $\Xi$.

 Let $\alpha \in (0,1)$, define $I_{\alpha,N}=\overline{1,[\alpha \cdot N]}$, where $[\cdot]$ is an
 integer part of the number. Now one can consider the real symmetric $N\times N$ matrix
$A^{(N,p,\alpha)}(\omega)$:
\begin{equation}\label{dilute}
\left[A^{(N,p,\alpha)}\right]_{ij}\!=\!\left\{ \begin{array}{ll} a_{ij}\cdot d_{ij}^{(N,p)},&
\textrm{if} \  (i \in I_{\alpha,N} \wedge j\notin I_{\alpha,N} ) \vee (i \notin I_{\alpha,N} \wedge
j\in I_{\alpha,N} )  ,
\\0,& \textrm{otherwise} \\ \end{array}
\right.
\end{equation}
that has $N$ real eigenvalues
$\lambda^{(N,p,\alpha)}_1\!\leq\!\lambda^{(N,p,\alpha)}_2 \!\leq\!\ \ldots
\!\leq\!\ \lambda^{(N,p,\alpha)}_N$.

The normalized eigenvalue counting function  of  $A^{(N,p,\alpha)}$ is determined
by the formula
$$
\sigma\left({\lambda; A^{(N,p,\alpha)}}\right)\!=\!\frac{\sharp
\{j:\lambda^{(N,p,\alpha)}_j\!<\!\lambda\}}{N}.
$$
 We denote by $\mathcal{F}$  the class
 of functions which are
 analytic with respect to $z:\,\,\Re z>0$ and for any fixed $z:\,\,\Re z>0$
 possessing the norm
 \begin{equation}
||f(u,z)||=\sup_{u>0}\frac{|f(u,z)|}{\sqrt{1+u}}.
\label{norm}\end{equation}

\begin{theorem}\label{thm:3}
Assume that $\mu(a)={\mathbb E}\{\theta(a-a_{i,j})\}$  the probability distribution
of $a_{i,j}$ possesses the property
\begin{equation}
\int a^4d\mu(a)=X_4<\infty. \label{cond_mu}\end{equation}
(We denote by $X_i$ i absolute moment of $a$, i.e. $X_i=\int |a|^id\mu(a)$.) Then
 the measure $\sigma\left({\lambda; A^{(N,p,\alpha)}}\right)$ converges weakly  in probability to nonrandom measure $\sigma_{p,\alpha}$
\begin{equation}
\sigma\left({\cdot \ ; A^{(N,p,\alpha)}}\right)\stackrel{w,P}{\to} \sigma_{p,\alpha},\ N \to\infty .
\end{equation}
  The Stieltjes transform $ g_{\sigma_{p,\alpha}}$ of the limiting measure  $\sigma_{p,\alpha}$
 can be found as follows:
 \begin{equation}
 g_{\sigma_{p,\alpha}}(z)=-ih(iz)
 \end{equation}
 \begin{equation}\label{h1}
 h(z):{\mathbb C}_{+}\to{\mathbb C}_{+}, \ h(z)=\alpha\cdot h_1(z)+(1-\alpha)\cdot h_2(z)
 \end{equation}
 \begin{equation}\label{h2}
 h_i(z)=-X_2^{-1}\fracd{\partial}{\partial
u}f_i(u,z)\bigg|_{u=0},\ i=1,2,
 \end{equation}
   where a pair  $f_1(u,z)$ and $f_2(u,z)$ is a unique solution of the following  system of functional equations in the class $\cal{F}$:
  \begin{equation} \label{system}
  \left\{ \begin{array}{l} f_1(u,z)=L(f_2,\mu,1-\alpha)
\\f_2(u,z)=L(f_1,\mu,\alpha) \end{array}
\right.,
   \end{equation}
   where
   \begin{equation}
  L(f,\mu,\alpha)=1-u^{1/2}e^{-\alpha p}\int|a| d\mu(a)\int_0^\infty dv
\frac{\mathcal{J}_1(2|a|\sqrt{uv})}{\sqrt v}
\exp\{-zv+\alpha p f(v,z)\},
   \end{equation}
   $\mathcal{J}_1(\zeta)$  is the Bessel function:
\begin{equation}
\mathcal{J}_1(\zeta)=\frac{\zeta}{2}\sum_{k=0}^{\infty}
\frac{(-\zeta^2/4)^k}{k!(k+1)!}, \label{J_1}\end{equation}
   \end{theorem}
\begin{proposition}\label{rem_1}
 Condition (\ref{cond_mu}) in Theorem \ref{thm:3} could be replaced by
 $$\int a^2d\mu(a)=X_2<\infty$$ via truncation procedure.
\end{proposition}
The proof of Proposition \ref{rem_1} is given in section \ref{section_4}.

Theorem \ref{thm:3} is a corollary of Theorem \ref{thm:4}.

 \begin{theorem}\label{thm:4}
 Let the distribution of $a_{j,k}$ satisfy condition
  (\ref{cond_mu}). Then

\noindent (i) the variance of  $g_{N,p,\alpha}(z)$
  vanishes in the limit $N\to\infty$:
\begin{equation}
{\mathbb E}\{|g_{N,p,\alpha}(z)-{\mathbb E}\{g_{N,p,\alpha}(z)\}|^2\}\le \fracd{C(z,p,\alpha,X_2)}{N^{1/2}},
\label{t3.2}
\end{equation}
\noindent (ii) there exists the limiting probability measure $\sigma_{p,\alpha}$ such that
\begin{equation}
g_{\sigma_{p,\alpha}}(z)=\lim_{N\to\infty}{\mathbb E}\{g_{N,p,\alpha}(z)\}=-ih(iz), \label{t3.4}
\end{equation}
  for arbitrary compact in $\mathbb{C}_{+}$ the convergence is  uniform, and the function $h(z):\mathbb{C}_{+}\to\mathbb{C}_{+}$ can be expressed in terms of the pair of functions $f_1(u,z)$ and $f_2(u,z)$ (see (\ref{h1})-(\ref{h2})), which is a unique solution of the  system of functional equations  (\ref{system}) in the class $\cal{F}$.
 \end{theorem}

\section{Proof of Theorem 1}

For any $z$: $\Re z>0$
consider the functions $f_{1,N}(u,z):{ \R}_+\to{\C}$, $f_{2,N}(u,z):{ \R}_+\to{\C}$:
\begin{equation}
f_{1,N}(u,z)=\frac{1}{[\alpha N]}\sum_{k=1}^{[\alpha N]}e^{-ua_k^2G_{kk}^{(N,p,\alpha)}(z)},\quad
G^{(N,p,\alpha)}(z)=(z-iA^{(N,p,\alpha)})^{-1},
\label{t3.1}\end{equation}
\begin{equation*}
f_{2,N}(u,z)=\frac{1}{N-[\alpha N]}\sum_{k=[\alpha N]+1}^Ne^{-ua_k^2G_{kk}^{(N,p,\alpha)}(z)},
\label{t3.1a}\end{equation*}
 where $\{a_i\}_{i=1}^\infty$  is a
family of i.i.d. random variables which do not depend on
  $\{a_{i,j}\}_{i<j}^\infty$  and have the same probability distribution as
  $a_{1,2}$.
 It is easy to see that $G_{NN}^{(N,p,\alpha)}(z)$ can be represented in the form
   \begin{equation} \label{alg}
G_{NN}^{(N,p)}(z)=\bigg(z -i A_{NN}^{(N,p,\alpha)} +
\sum_{j,k=1}^{N-1} \widetilde
G_{jk}^{(N-1,p,\alpha)}A_{Nj}^{(N,p,\alpha)}A_{Nk}^{(N,p,\alpha)}\bigg)^{-1},
\end{equation}
where the matrix $\{\widetilde G_{ij}^{(N-1,p,\alpha)}(z)\}_{i,j=2}^N$ is
a resolvent of the matrix $i\widetilde A^{(N-1,p,\alpha)}$, which can be
obtained from $A^{(N,p,\alpha)}$ by deleting the last column and the last row. Let us use the formula
  (see \cite{AS}):
\begin{equation} \label{t3.5a}
e^{-ua^2R}=1-u^{1/2}|a|\int_0^\infty dv
\frac{\mathcal{J}_1(2|a|\sqrt{uv})}{\sqrt v} \exp\{-R^{-1}v\},
  \end{equation}
which is valid for any $u\ge 0$, $\Re R>0$. Then, on the basis of
(\ref{alg}), we get
\begin{equation}\begin{array}{lcl}
\exp\{-ua_N^2G_{NN}^{(N,p,\alpha)}\} &=&1-u^{1/2} |a_N|\intd_0^\infty dv
\fracd{\mathcal{J}_1(2|a_N|\sqrt{uv})}{\sqrt v}\\
&&\exp\{-zv-v\sumd_{j,k=1}^{[\alpha N]}  \widetilde G_{ij}^{(N-1,p,\alpha)}
A^{(N,p,\alpha)}_{Ni}A^{(N,p,\alpha)}_{Nj}\}.
\end{array}\label{t3.7}\end{equation}
Denote
\begin{equation}  R_N(z)=\sum_{j,k=1,j\not=k}^{[\alpha N]}
  \widetilde G_{jk}^{(N-1,p)}A_{Nj}^{(N,p,\alpha)}A_{Nk}^{(N,p,\alpha)}.
 \label{t3.R_N}\end{equation}

\begin{proposition}\label{pro_1}
\begin{equation}
{\E}\{|R_1(z)|^2\}\le 2\frac{p^2 X_2^2}{N|\Re z|^2}+
\frac{p^4X_1^4}{N^2|\Re z|^2}+
4\frac{p^3 X_1^2X_2}{N^2|\Re z|^2}+
6\frac{p^4X_1^4}{N|\Re z|^2}+
8\frac{p^3 X_1^2X_2}{N|\Re z|^2}, \label{t3.6}\end{equation}
\end{proposition}
where
$$
X_k=\int |a|^kd\mu(a).
$$
Proof.
\begin{equation}\begin{array}{rcl}
{\E}\{|R_1(z)|^2\}&=&\sumd_{\not=\{j_1,j_2,k_1,k_2\}}^{[\alpha N]}\E\bigg\{
\widetilde G_{j_1k_1}^{(N-1,p)}\overline{\widetilde
  G_{j_2k_1}^{(N-1,p)}}
A^{(N,p)}_{Nj_1}A^{(N,p)}_{Nj_2}A^{(N,p)}_{Nk_1}A^{(N,p)}_{Nk_2}\bigg\}\\
&&+4\sumd_{\not=\{j,k_1,k_2\}}^{[\alpha N]} {\E}\bigg\{\widetilde
G_{jk_1}^{(N-1,p)}\overline{\widetilde
  G_{jk_1}^{(N-1,p)}}
|A^{(N,p)}_{Nj}|^2A^{(N,p)}_{Nk_1}A^{(N,p)}_{Nk_2}\bigg\}\\
&&+2\sumd_{j\not=k}^{[\alpha N]}{\E}\bigg\{ \widetilde G_{jk}^{(N-1,p)}
\overline{ \widetilde G_{jk}^{(N-1,p)}}
|A^{(N,p)}_{Nj}|^2|A^{(N,p)}_{Nk}|^2\bigg\}=H_1+4H_2+2H_3.
\end{array}\label{t3.6a}\end{equation}
Averaging with respect to $\{A^{(N,p)}_{N,i}\}_{i=1}^{N-1}$ and using
the fact that $\{\widetilde G_{ij}^{(N-1,p)}(z)\}_{i,j=1}^{N-1}$ do not
depend on
  $A^{(N,p)}_{N,i}$, we obtain
$$\begin{array}{rcl}
H_1&\le&X_1^4\fracd{p^2}{N^2}{\E}\bigg\{\bigg|N^{-1}\sumd_{j,k}^{[\alpha N]}{\widehat G}_{jk}\bigg|^2\bigg\}+
6\frac{p^4X_1^4}{N|\Re z|^2}\le\fracd{p^2 X_1^4}{N^2|\Re z|^2}+
6\frac{p^4X_1^4}{N|\Re z|^2}\\
H_2&\le & X_1^2X_2\fracd{p}{N^3}\sumd_{k_1\not=k_2}^{[\alpha N]}{\E}\bigg\{[{\widehat G}{\widehat G}^{*}]_{k_1
k_2}\bigg\}+
2\frac{p^3 X_1^2X_2}{N|\Re z|^2}\le \fracd{p X_1^2X_2}{N^2|\Re z|^2}+
2\frac{p^3 X_1^2X_2}{N|\Re z|^2},\\
H_3&\le &\fracd{X_2^2}{N^2}\sumd_{k}^{[\alpha N]}{\E}\bigg\{[{\widehat G}{\widehat G}^{*}]_{kk}\bigg\}\le \fracd{X_2^2}{N|\Re z|^2}.
\end{array}$$
Besides, since evidently
$$
\Re\bigg\{\sumd  \widetilde G_{ij}^{(N-1,p)}
A^{(N,p,\alpha)}_{Ni}A^{(N,p,\alpha)}_{Nj}\bigg\}\ge 0,\quad \Re\bigg\{\sumd  \widetilde
G_{jj}^{(N-1,p,\alpha)} |A^{(N,p,\alpha)}_{jj}|^2\bigg\}\ge 0,
$$
the inequality
\begin{equation}\label{exp_ineq}
|e^{-z_1}-e^{-z_2}|\le |z_1-z_2| \quad (
\Re z_1,\Re z_2\ge 0)
  \end{equation}
   and (\ref{t3.7}) imply
  \begin{equation}\begin{array}{lcl}
\E\exp\{-ua_N^2G_{NN}^{(N,p,\alpha)}\} &=&1-u^{1/2}\E |a_N|\intd_0^\infty dv
\fracd{\mathcal{J}_1(2|a_N|\sqrt{uv})}{\sqrt v}\\
&& \E _2\exp\{-zv-v\sumd_{i=1}^{[\alpha N]}  \widetilde G_{ii}^{(N-1,p,\alpha)}
|A^{(N,p,\alpha)}_{Nj}|^2\}+\widetilde r_N(u),
\end{array}\label{t3.8}\end{equation}
where $\E_2$ denotes averaging over $\left\{a_{ij}\right\}_{i,j}$ and $\left\{d^{(N,p)}_{ij}\right\}_{i,j}$. Remainder $\widetilde r_N(u)$ obeys the following condition
$$
|\widetilde r_N(u)|\le \E | R_1|u^{1/2}\E |a_N|\intd_0^\infty dv
|\fracd{\mathcal{J}_1(2|a_1|\sqrt{uv})}{\sqrt v}e^{-zv}|\le
C| R_1|u^{1/2}\E |a_N||\Re z|^{-1/2}.
$$
In the last inequality we use the estimate for the Bessel function
\begin{equation}\label{Bessel}
|\mathcal{J}_1(u)|\le 1.
\end{equation}

Here and below we denote by  $C$ some constants (different in
different formulas), which do not depend on  $N,z,p,\alpha$. Taking into
account  (\ref{t3.6}), we get
\begin{equation}
|\widetilde r_N(u)| \le\frac{C(p,X_2)u^{1/2}}{\sqrt{N}|\Re z|^{5/2}}.
\label{t3.8a}\end{equation}

Since
$\widetilde G^{(N-1,p)}(z)$ does not depend on
  $\left\{d^{(N,p)}_{N,j}\right\}_{j=1}^N$, we obtain
\begin{equation}\begin{array}{rcl}
\E _2\exp\{-v\sumd_{i=1}^{[\alpha N]}  \widetilde G_{ii}^{(N-1,p,\alpha)}
|A^{(N,p,\alpha)}_{Nj}|^2\}&=&\E \bigg\{\prodd_{j=1}^{[\alpha N]}
\bigg((1-\fracd{p}{N})+\fracd{p}{N}e^{-va_{Nj}^2\widetilde
G_{jj}^{(N-1,p)}}\bigg)
\bigg\}\\
&=&e^{-\alpha p}\E \bigg\{ \exp\{\alpha p\widetilde f_{1,N-1}(v,z)\}\bigg\}
+ R_{N}(v),
\end{array}\label{t3.10}\end{equation}
where
$$
\widetilde f_{1,N-1}(u,z)=\frac{1}{[\alpha N]}\sum_{k=1}^{[\alpha N]}e^{-va_{Nj}^2\widetilde G_{kk}^{(N,p,\alpha)}(z)},
$$
and $R_{N}$ satisfy the estimate
$$
|R_{N}(v)|\le\fracd{Cp^2}{N}.
$$
Using  (\ref{exp_ineq}), we get
\begin{multline}
\left| \E \exp\left\{\alpha p  \widetilde f_{1,N-1}(v,z)\right\}-
    \exp\left\{\alpha p \E\widetilde f_{1,N-1}(v,z)\right\}\right|\le
    \\\label{use_ineq}
\le \alpha p e^{\alpha p}\E \left|\widetilde f_{1,N-1}(v,z)-\E \widetilde
f_{1,N-1}(v,z)\right|
\end{multline}

 Further considerations are based on the Lemma \ref{martingal}.

\begin{lemma}\label{martingal}
 Fix $\alpha \in (0,1)$. Let $A^{(n)}$ be a  real symmetric n$\times$ n matrix, such that
 \begin{equation}
A^{(n)}_{ij}\!=\!\left\{ \begin{array}{ll} a_{ij},&
\textrm{if} \  (i \in I_{\alpha,n} \wedge j\notin I_{\alpha,n} ) \vee (i \notin I_{\alpha,n} \wedge
j\in I_{\alpha,n} )  ,
\\0,& \textrm{otherwise} \\ \end{array}
\right.
\end{equation}
where $\{a_{ij}\}_{1\le i\le j}$ is a family of
jointly independent identically distributed  random
variables that obey the following conditions
\begin{equation}\E |a_{ij}|\le \frac{C}{n}, \ \E a_{ij}^2\le \frac{C}{n}
\end{equation}
\
 For  $z$: $\Re z>0$  consider
\begin{equation}\label{F(R)}
  R=(z-iA)^{-1},\quad F_n(z)=[\alpha\cdot n]^{-1}\sum_{j=1}^{[\alpha\cdot n]} \varphi_j(R_{jj}),
\end{equation}
 where random functions $\varphi_j$ satisfy the following condition
\begin{equation}|\varphi'_j(\zeta)|\le C_3\cdot \alpha_j.
\end{equation}
where $\left\{\alpha_j\right\}$ is a set of jointly independent identically distributed  random
variables  also independent of $\left\{a_{ij}\right\}$ such that $\E\left\{\alpha_1^2\right\}<\infty$.
Then
\begin{equation}\label{variation}
Var F_n(z)\le  \fracd{4(1-\alpha)  C_3^2}{\alpha n |\Re z|^4} \cdot \E\left\{\alpha_1^2\right\}\cdot
\left(n \E |a_{12}|+ (n \E a^2_{12})^2\right).
\end{equation}
\end{lemma}
The proof of Lemma \ref{martingal} is given in section \ref{section_4}.

\begin{remark}
Lemma \ref{martingal} is still valid for $F_n(z)= n^{-1}\sum_{j=1}^{ n} \varphi_j(R_{jj})$ with changed constants.
\end{remark}

  Lemma \ref{martingal} for
 $\varphi(\zeta)=\exp{\{-va_{Nj}^2\zeta\}}$, $\alpha_j=va_{Nj}^2$, $C_3=1$, $n=N-1$ implies

\begin{equation}\label{var_tilda_f}
\E\left|\widetilde  f_{1,N-1}(v,z)-\E \widetilde f_{1,N-1}(v,z)\right|^2
\le\fracd{\widetilde{C}^2(X_4,p) v^2}{\alpha N \left|\Re
z\right|^4}
\end{equation}
Relations (\ref{use_ineq}) and (\ref{var_tilda_f}) yield
\begin{equation}  \label{asc_E_tilda}
\left| \E \exp\left\{\alpha p  \widetilde f_{1,N-1}(v,z)\right\}-
    \exp\left\{\alpha p \E\widetilde f_{1,N-1}(v,z)\right\}\right|
   \le \alpha p e^{\alpha p} \fracd{\widetilde{C}(X_4,p) v}{\alpha^{1/2} N^{1/2} \left|\Re
z\right|^2}
\end{equation}
Combining (\ref{t3.7})-(\ref{asc_E_tilda}), we get
\begin{equation}\label{recur_eq_tilda} \E
f_{2, N}(u,z)=1-u^{1/2}e^{-\alpha p}\intd_0^\infty dv
e^{-zv}\fracd{\mathcal{J}_1(2\sqrt{uv})}{\sqrt
v}e^{\alpha p \E \widetilde f_{1,N-1}(v,z)}+r(u),
\end{equation}
$$
 r(u)\le\fracd{\widetilde{C}(\E a^4_{Nj},p) u^{1/2}}{N^{1/2}\left|\Re
z\right|^{7/2}}.
$$
In order to get the closed system of equations, we have to replace $\widetilde f_{1,N-1}$ by $f_{1, N}$.
 To this purpose we use the next bound on their difference
\begin{equation}\label{tilda}
\left|\E f_{1,N}(v,z)-\E\widetilde  f_{1,N-1}(v,z)\right|\le
\fracd{\widetilde{C}(X_4,p) v}{\alpha N^{1/2} \left|\Re
z\right|^2}.
\end{equation}
  Indeed, using Lemma \ref{martingal} for
 $\varphi(\zeta)=\exp{\{-va_{j}^2\zeta\}}$, $\alpha_j=va_{j}^2$, $C_3=1$, $n=N$,
we get
\begin{equation}\label{var_f_for_prop}
\E\left| f_{1,N}(v,z)-\E  f_{1,N}(v,z)\right|^2
\le\fracd{\widetilde{C}^2(X_4,p) v^2}{\alpha N \left|\Re
z\right|^4}.
\end{equation}
 Combining (\ref{var_f_for_prop}), (\ref{var_tilda_f}) and (\ref{qq5}) for
  $\varphi(\zeta)=\exp{\{-va_{Nj}^2\zeta\}}$, $\alpha_j=va_{Nj}^2$, $C_3=1$, $n=N$,
  we obtain (\ref{tilda}).

  The inequalities (\ref{tilda}), (\ref{exp_ineq}) and (\ref{recur_eq_tilda})  imply
\begin{equation}\label{recur_eq} \E
f_{2, N}(u,z)=1-u^{1/2}e^{-\alpha p}\intd_0^\infty dv
e^{-zv}\fracd{\mathcal{J}_1(2\sqrt{uv})}{\sqrt
v}e^{\alpha p \E f_{1,N}(v,z)}+r(u),
\end{equation}
$$
 r(u)\le\fracd{\widetilde{C}(X_4,p) u^{1/2}}{N^{1/2}\left|\Re
z\right|^{7/2}}.
$$
Let us  consider the Banach space ${\cal
H}$ of all the functions $h:\mathbb{R}_+\to \mathbb{C}$ which
possess the norm (\ref{norm}). ${\cal H}^2$ possess the norm
$
||(h_1,h_2)||_{{\cal
H}^2}=\max\left\{||h_1||_{\cal H},||h_2||_{\cal H}\right\}.
$

Define the operator  $F_z:{\cal H}^2\to{\cal H}^2$ of the form
\begin{equation}
F_z(\varphi_1,\varphi_2) = (\psi_1,\psi_2), \
\psi_1(u)=L(f_2,\mu,1-\alpha), \
\psi_2(u)=L(f_1,\mu,\alpha).
\label{operator}
\end{equation}
Let us  denote  by $B_{0,2}=\{h\in\mathcal{H}^2,\|h\|_{\mathcal{H}^2}\le 2\}$ the
ball of radius  2 centered in the origin.
 Then for any  $\varphi_1$,
$\varphi_2:$  $||\varphi_{1}||\le 2$, $\|\varphi_{2}\|\le 2$
\begin{equation}\label{vicinity1}
\|F_z(\varphi_1)-F_z(\varphi_2)\|\le X_1 p e^{2p+\frac{2p^2}{|\Re
z|}}\left(\fracd{2}{|\Re z|}+\fracd{4}{|\Re z|^{1/2}}\right)
 \|\varphi_1-\varphi_2\|.
 \end{equation}

 Indeed, inequalities (\ref{exp_ineq}), (\ref{Bessel}) imply
 \begin{equation}\label{vicinity1a}
 \|F_z(\varphi_1)-F_z(\varphi_2)\|_{\mathcal{H}^2}\le X_1 p \|\varphi_1-\varphi_2\| \int_0^\infty
\frac{dv(1+v^{1/2})e^{-|\Re z|v+2p(1+v^{1/2})}}{\sqrt v}
  \end{equation}
  Using the trivial inequality
  \begin{equation}\label{vicinity1b}
  2pv^{1/2}-v|\Re z|/2\le2\fracd{p^2}{|\Re z|},
  \end{equation}
  we get (\ref{vicinity1}).

  It is easy to see that  $\|F_z(0)\|$ obey the inequality
  \begin{equation}\label{norm_0}
 \|F_z(0)\|_{\mathcal{H}^2}\le  \supd_{u>0} \fracd{1+2X_1|\Re z|^{-1/2}u^{1/2}}{1+u^{1/2}}\le 1+2X_1|\Re z|^{-1/2}.
  \end{equation}

  So there is  $M>0$ such that

\begin{equation}\label{vicinity2}
\|F_z(\varphi_1)-F_z(\varphi_2)\|< 1/4
 \|\varphi_1-\varphi_2\|,\quad \|F_z(0)\|\le \fracd{5}{4},\quad z\in L(M),
 \end{equation}
where $L(M)=\{z:|\Re z|>M\}.$

 Therefore
 $F_z:B_{0,2} \to B_{0,2}$ and  $F_z|_{B_{0,2}}$ is a contraction mapping for all  $ z\in L(M)$. Hence, there exists the unique fixed point $f(u,z)$,
which is  a solution of (\ref{system}).
Since $|\E f_{1,N}(u,z)|\le 1, |\E f_{2,N}(u,z)|\le 1,$  $\E f_N(u,z)\in B_{0,2}$. (\ref{vicinity2}) imply the estimates
\begin{equation}
\E f_N(u,z)=F_z(\E f_N(u,z))+r_N(u,z)=\ldots =f(u,z)+r^{\infty}_N(u,z),
\end{equation}
where
\begin{equation}
\|r^{\infty}_N(u,z)\|\le \|r_N(u,z)\|\sumd_{k=0}^{\infty}\fracd{1}{4^k}\le \fracd{4C(p,M)}{3N^{1/2}}, \quad z\in L(M).
\end{equation}

Hence, $\E f_{N}(u,z)\rightrightarrows f(u,z),\quad z\in L(M)$. Fix u. Since $\E f_{1,N}(u,z)$, $\E f_{2,N}(u,z)$ are analytic and uniformly bounded for arbitrary $\Pi_{\varepsilon,a}=\{z:\varepsilon\le \Re z \le 2M, |\Im z| \le a \}$,  by the  Arzela theorem we can choose a subsequence
$\{N_k\}_{k=1}^{\infty}$ such that $\E f_{1,N_k}(u,z)\rightrightarrows \widetilde f_1^{(a,\varepsilon)}(u,z)$, $\E f_{2,N_k}(u,z)\rightrightarrows \widetilde f_2^{(a,\varepsilon)}(u,z)$ in $\Pi_{\varepsilon,a}$.  Then $f_1^{(a,\varepsilon)}(u,z)$,
$f_2^{(a,\varepsilon)}(u,z)$ are analytic in $\Pi_{\varepsilon,a}$. But
$$
f_2^{(a,\varepsilon)}(u,z)=f_2(u,z), \ f_1^{(a,\varepsilon)}(u,z)=f_1(u,z),\quad  |\Re z|>M.
$$

 The uniqueness theorem of complex
analysis and an arbitrariness of choosing subsequence imply the existence of the analytic extension of
$f_1(u,z)$ ($f_2(u,z$)  in $\C_{+}$ and uniform convergence in
$z$ $f_{\alpha,N}(u,z)\rightrightarrows f_{\alpha}(u,z)$ for arbitrary compact in $\C_{+}$. Thus, if we fixed any $z:\Re z>0$, we obtain that $f_{\alpha,N}(u,z), (\alpha=1,2)$ as a function of $u$ converges pointwise to $f_{\alpha,N}(u,z)$. But since $|\frac{\partial}{\partial
u}f_{\alpha,N}(u,z)|\le C$ and $\left| f_{\alpha,N}(u,z)\right|\le 1$, the pointwise convergence imply also   the convergence in the norm (\ref{norm}). Then, using  Lebesgue's dominated convergence theorem, we prove that $f(u,z)=F_z(f)(u,z)$ in $\C_{+}$.

Indeed, using Lemma \ref{martingal} for
 $\varphi(\zeta)=\zeta$, $\alpha_j=1$, $C_3=1$, $n=N$,
we get (\ref{t3.2}).

Uniform convergence $f_N(u,z)$ in $u\in [0,1]$ imply
\begin{multline}
\E g_{N,p,\alpha}=-i(\E a_1^2)^{-1}\fracd{1}{N}\sumd_{k=1}^N \E a_k^2\E \{G_{kk}^{(N,p,\alpha)}\}=
\\=
i(\E a_1^2)^{-1}\left(\alpha\fracd{\partial}{\partial
u}\E f_{1,N}(u,z)\bigg|_{u=0}+(1-\alpha)\fracd{\partial}{\partial
u}\E f_{2,N}(u,z)\bigg|_{u=0}\right),
\end{multline}
The next simple proposition allows us to make a final step.
\begin{proposition}
Set $\Psi_n(u)=\fracd{f_n(u)-f_n(0)}{u}-f_n'(0)$. Assume that
\begin{equation}\label{*}
|\Psi_n(u)|\le\varepsilon(u),\; \varepsilon(u)\to 0, \; as \  u\to 0.
\end{equation}
 If there exists  $f(u)=\limd_{n \to \infty}f_n(u)$ and the function $f$ is differential at $u=0$, then
 \begin{equation}
 \limd_{n \to \infty}f_n'(0)=f'(0).
 \end{equation}
\end{proposition}
Proof.
  \begin{multline}
\Big|\fracd{f_{1,n}(u,z)-f_{1,n}(0,z)}{u}-\fracd{\partial}{\partial u}f_{1,n}(u,z)\bigg|_{u=0}\Big|=
\\
=\bigg|\int_0^1\left(\fracd{\partial}{\partial u}f_{1,n}(tu,z) -\fracd{\partial}{\partial u}f_{1,n}(u,z)\bigg|_{u=0}\right)dt \bigg|\le
 \\
\le \frac{1}{[\alpha N]}\sum_{k=1}^{[\alpha N]}\intd_0^1 a_k^2 \Big|G_{kk}^{(N,p,\alpha)}\Big|\bigg|e^{-uta_k^2G_{kk}^{(N,p,\alpha)}}-1\bigg|dtd\mu(a_k)\le
 \\
\le \fracd{1}{|\Re z|}\frac{1}{[\alpha N]}\sum_{k=1}^{[\alpha N]}\left(\intd_{|a^2|>u^{-1/2}} +\intd_{|a^2|\le u^{-1/2}}\right)\intd a^2\bigg|e^{-uta_k^2G_{kk}^{(N,p,\alpha)}}-1\bigg|d\mu(a_k)dt\le
 \\\
 \le\fracd{1}{|\Re z|}\left(2 \intd_{|a^2|>u^{-1/2}}a^2d\mu(a) +\fracd{1}{|\Re z|}\sqrt{u} \intd a^2d\mu(a)\right)\xrightarrow{u \to 0} 0.
  \end{multline}
  Similar estimate  with $f_{2,n}$ is valid too.
Hence,
\begin{equation}
g_{p,\alpha}(z)=\limd_{N\to\infty}\E g_{N,p,\alpha}(z)=i(\E a_1^2)^{-1}\left(\alpha\fracd{\partial}{\partial
u} f_{1}(u,z)\bigg|_{u=0}+(1-\alpha)\fracd{\partial}{\partial
u} f_{2}(u,z)\bigg|_{u=0}\right)
\end{equation}

\section{Proofs of auxiliary statements}\label{section_4}

\bigskip{\it Proof of Lemma \ref{martingal}.}

 Let us denote by  $\E_k$ averaging over  $\{a_{ij}\}_{i\le j}$
 $i\le k$
($\E_n=\E$, $\E_0$ means absence of averaging), by $n_1=[\alpha\cdot n]$, by $n_2=n-
n_1$. Then
\begin{multline}
 F_n-\E F_n=\sumd_{k=0}^{n_1-1}(
\E _kF_n-\E _{k+1}F_n)\Rightarrow \qquad \E |F_n-\E F_n|^2=
\\
=2\sumd_{j<k}^{n_1-1}\E( \E_k F_n-\E_{k+1} F_n) ( \E_j\overline
F_n-\E_{j+1}\overline F_n)+\sumd_{k=0}^{n_1-1} \E| \E_k F_n-\E_{k+1}F_n|^2=
\\
 =\sumd_{k=0}^{n_1-1}\E|
\E_k F_n-\E_{k+1}F_n|^2.\label{qq1}
\end{multline}
 Here we use the identity   $\E( \E_k F_n-\E_{k+1} F_n) (
\E_j\overline F_n-\E_{j+1}\overline F_n)=0$ for $j\neq k$.

Denote  by $\E^{(k)}$ averaging over
$\left\{a_{kj}\right\}_{j=k}^n$. Let $F_n^{(k)}=F_n\bigg|_{\left\{a_{kj}=0\right\}_{j=1}^n}$. So we get
$$
 \E| \E_k
F_n-\E_{k+1}F_n|^2=\E| \E_k (F_n-\E^{(k+1)}F_n)|^2\le \E\left| \E_k\left(
F_n-F_n^{(k+1)} \right)\right|^2.
$$
 Taking into account the  Schwarz inequality, we obtain $\E\left|
\E_k\left( F_n-F_n^{(k)} \right)\right|^2\le \E\E_k\left| \left(
F_n-F_n^{(k+1)} \right)\right|^2 = \E\left| F_n-F_n^{(k+1)} \right|^2$.
  Due to the symmetry we have for all $k$
\begin{equation}\label{qq2}
\E| \E_k F_n-\E_{k+1}F_n|^2\le \E\left| F_n-F_n^{(k+1)} \right|^2=\E\left|
F_n-F_n^{(1)}\right|^2
\end{equation}
  In order to estimate  $\E\bigg| F_n-F_n^{(1)}\bigg|^2$, we introduce the matrix $A(t)$  : $$\begin{array}{l} A_{ij}(t)= \left(\1_{i\in  I_{\alpha,N} , j\notin I_{\alpha,N} }+\1_{i\notin  I_{\alpha,N} , j\in I_{\alpha,N} }\right)\cdot\left(\1_{i\ge 1, j\ge 1}a_{ij}+ t\cdot(1-\1_{i\ge 1, j\ge 1})\cdot a_{ij}\right) .
\end{array}$$
Also we introduce the functions $$R(t)=(z-iA(t))^{-1},\quad F_n(t)=n_1^{-1}\sum_{j=1}^{n_1}
\varphi(R_{jj}(t)). $$ Clearly, the following equality is true
\begin{equation}\label{l1.3}
F_n-F_n^{(1)}=\int_0^1\frac{d}{dt}F_n(t) dt.
\end{equation}
We can estimate $\fracd{d}{dt}F(t)$ by the following way:
\begin{equation*}
\fracd{d}{dt}F_n(t)=
\fracd{1}{n_1}\sumd_{j=1}^{n_1}\sumd_{k,l}\xi_j(t)R_{jk}(t)A_{kl}'(t)R_{lj}(t)=\fracd{2}{n_1}\sumd_{j\in I_{\alpha,n},k,l}\xi_j(t)R_{j1}(t)a_{1k}(t)R_{kj}(t)=2H,
\end{equation*}
where $$\xi_j(t)=\frac{\partial}{\partial
R_{jj}}\varphi(R_{jj}(t)).$$
\begin{multline}
\E\left\{\left|H\right|^2\right\}\le \fracd{C_3^2}{4n_1^2}\E \left\{\sumd_ {j,j'\in I_{\alpha,n},k,k'} \alpha_j \alpha_j'\left( \left|R_{jk}\right|^2+\left|R_{1j}\right|^2\right)|a_{1k}|\left( \left|R_{j'k'}\right|^2+\left|R_{1j'}\right|^2\right)|a_{1k'}|\right\}\le
\\
\le\fracd{C_3^2}{n_1^2}\E\left\{\alpha_1^2\right\} \E\left( \sumd_k|a_{1k}|^2\right)\le
\fracd{C_3^2}{n_1^2|\Re z|^4}\E\left\{\alpha_1^2\right\} \cdot
\left(n_2 \E |a_{12}|+ (n_2 \E a^2_{12})^2\right)\label{qq3}
\end{multline}
Here we use the inequality
$$
\sumd_j \left|R_{sj}\right|^2=\left[RR^*\right]_{ss}\le\fracd{1}{|\Re z|^2}.
$$

 Using (\ref{l1.3}) and (\ref{qq3}), we obtain
\begin{multline}
 \E\left|F_n-F_n^{(1)}\right|^2=\E \left\{\left|\intd_0^1\fracd{d}{dt}F_n(t)\right|^2\right\}\le
 \intd_0^1 \E \left\{\left|\fracd{d}{dt}F_n(t)\right|^2\right\}\le
 \\
 \le \fracd{4  C_3^2}{ n_1^2 |\Re z|^4} \cdot \E\left\{\alpha_1^2\right\}\cdot
\left(n_2 \E |a_{12}|+ (n_2 \E a^2_{12})^2\right).
  \label{qq4}
  \end{multline}
  In much  the same way  the following estimate can be proved
  \begin{multline}
 \E\left|F_n-F_n^{(n)}\right|^2=\E \left\{\left|\intd_0^1\fracd{d}{dt}F_n(t)\right|^2\right\}\le
 \intd_0^1 \E \left\{\left|\fracd{d}{dt}F_n(t)\right|^2\right\}\le
 \\
 \le \fracd{4  C_3^2}{ n_1^2 |\Re z|^4} \cdot \E\left\{\alpha_1^2\right\}\cdot
\left(n_2 \E |a_{12}|+ (n_2 \E a^2_{12})^2\right).
  \label{qq5}
  \end{multline}
  Combining (\ref{qq1}), (\ref{qq2}),  (\ref{qq4}) we obtain  (\ref{variation}).

\bigskip{\it Proof of Proposition \ref{rem_1}.}
 Denote by $a^{(T)}$ truncation of $a$ with parameter $T$, i.e.
\begin{equation}
a^{(T)}(\omega)\!=\! \left\{ \begin{array}{ll} a(\omega),&
\textrm{if} \  a(\omega)<T,
\\T,& \textrm{otherwise} \\ \end{array}
\right. .
\end{equation}

  Here and below the notation with an upper index $(T)$   means that the function is defined for the matrix $A^{(T)}$ by the same way as it was done for $A$.

 Similar to (\ref{vicinity1a}) and (\ref{vicinity1b}), we can obtain that
  for any  $\varphi$:
 $||\varphi||\le 2$
\begin{equation}\label{first_step}
\|F_z(\varphi)-F_z^{(T)}(\varphi)\|\le\fracd{2e^{2p+\frac{2p^2}{|\Re
z|}}}{|\Re z|}\|\varphi\|\int_{T}^{\infty}|a|d\mu(a)
 .
 \end{equation}
 Combining (\ref{first_step}) and (\ref{vicinity2}), we  obtain
\begin{equation}\label{second_step}
\forall z \in L(M) \ f^{(T)}(u,z)\xrightarrow[T\to \infty]{\|\|_{{\cal H}^2}} f(u,z)
 .
 \end{equation}
  Theorem \ref{thm:4} yields

\begin{equation}\label{third_step}
g_{\sigma^{(T)}_{p,\alpha}}(z)=\lim_{N\to\infty}{\mathbb E}\{g^{(T)}_{N,p,\alpha}(z)\}, \end{equation}
 and for an arbitrary compact in $\mathbb{C}_{+}$ the convergence is  uniform.

Taking into account the resolvent identity and the  Schwarz inequality, we obtain
 \begin{equation}\label{forth_step}
\Big|\E g_{\sigma^{(T)}_{N,p,\alpha}}(z)-\E g_{\sigma_{N,p,\alpha}}(z)\Big| \le \fracd{p}{N|\Im z|^2}\left(\int_{T}^{\infty}a^2d\mu(a)\right)^{1/2}
\end{equation}
Combining (\ref{second_step})-(\ref{forth_step}), we  obtain
\begin{equation}g_{\sigma_{p,\alpha}}(z)=\lim_{N\to\infty}{\mathbb E}\{g_{N,p,\alpha}(z)\}, \end{equation}
 and for an arbitrary compact in $\mathbb{C}_{+}$ the convergence is  uniform.

 (\ref{t3.2}) is still valid because it doesn't require existence of $X_4$ (Just use Lemma \ref{martingal} for
 $\varphi(\zeta)=\zeta$, $\alpha_j=1$, $C_3=1$, $n=N$).


\begin{thebibliography}{99}




\bibitem{AS} M.Abramowitz, I.Stegun, \emph{Handbook of Mathematical
Functions}, Dover, N.Y., 1972




\bibitem{BG1} M.Bauer and O.Golinelli. Random incidence matrices:
spectral density at zero energy, Saclay preprint T00/087;
cond-mat/0006472



\bibitem{B} B. Bollobas {\it Random Graphs }   Acad. Press (1985)


\bibitem{BG2} M.Bauer and O.Golinelli. Random incidedence matrices:
moments and
spectral density, J.Stat. Phys. {\bf 103}, 301-336, 2001



\bibitem{Ch} Fan R.K. Chung, {\it Spectral Graph Theory}  {\bf }
AMS (1997)


\bibitem{CDS}  D.M. Cvetkovi$ \acute{c}$, M.Doob, and H.Sachs.
        \textit{Spectra of Graphs}, Acad. Press (1980)



\bibitem{E1} S.N. Evangelou. Quantum percolation and the Anderson
transition in dilute systems, {\it Phys. Rev. B }  {\bf 27} (1983)
1397-1400


\bibitem{E2} S.N. Evangelou and E.N. Economou. Spectral density
singularities, level statistics, and localization in
sparse random matrices,
  {\it Phys. Rev. Lett. }  {\bf 68} (1992) 361-364

\bibitem{E3} S.N. Evangelou. A numerical study of sparse
random matrices,  {\it J. Stat. Phys. }  {\bf   69}
(1992) 361-383

\bibitem{FM:96} Y.V.Fyodorov, A.D.Mirlin. Strong eigenfunction
correlations near the Anderson localization transition,  {\it Phys. Rev. B }  {\bf 55} (1997)
R16001--R16004





\bibitem{GR} Ch. Godzil, G. Royle, {\it  Algebraic Graph Theory.}  {\bf }
Springer-Verlag, New York (2001)


\bibitem{JLR}  S. Janson, T. \L uczak, A. Rucinski, {\it Random Graphs. }
{\bf }  John Wiley \& Sons, Inc. New York (2000)

\bibitem{JMRR} D. Jacobson, S.D. Miller, I. Rivin,
and Z. Rudnick. Eigenvalue spacing for regular graphs,
in: {\it Emerging applications of number theory. } Ed. D.A. Hejhal et al.
{\bf } Springer-Verlag  (1999)








\bibitem{KSV} Khorunzhy O., Shcherbina M., and Vengerovsky V. Eigenvalue distribution of large
    weighted random graphs, J. Math. Phys. {\bf 45}  N.4: (2004), 1648-1672.


\bibitem{KKPS}  A. Khorunzhy, B. Khoruzhenko, L. Pastur and
M. Shcherbina. The Large-n Limit in Statistical Mechanics and
Spectral Theory
     of Disordered Systems.
     Phase transition and critical phenomena.v.15, p.73, Academic
     Press, 1992
\bibitem{Me:91} M.L.Mehta: \emph{Random Matrices}. New York: Academic
Press, 1991



\bibitem{MF:91} A.D.Mirlin, Y.V.Fyodorov. Universality of the
level correlation function of sparce random matrices,
  J.Phys.A:Math.Jen.{\bf 24}, (1991), 2273-2286.



\bibitem{RB:88} G.J. Rodgers  and A.J. Bray. Density of states of a
sparse random matrix, Phys.Rev.B {\bf 37}, (1988), 3557-3562.

\bibitem{RD:90}  G.J. Rodgers and C. De Dominicis. Density of states of
sparse random matrices,
J.Phys.A:Math.Jen.{\bf 23}, (1990), 1567-1566.

\bibitem{V} V. Vengerovsky. Eigenvalue Distribution of a Large Weighted Bipartite Random Graph,
JMPAG {\bf 10-2} (2014), 240-255.

\bibitem{W} E.P.Wigner. On the distribution of the roots of
certain symmetric matrices, Ann.Math. {\bf 67}: (1958), 325-327.


\end{thebibliography}
\end{document}